\documentclass[a4paper]{jpconf}
\usepackage{graphicx,amsmath,amsfonts,cite}

\def\id{\textrm{id}}
\newcommand{\bmat}[1]{\begin{bmatrix}#1\end{bmatrix}} 
\newcommand{\ket}[1]{\left|#1\right\rangle}
\newcommand{\bra}[1]{\langle#1|}
\newcommand{\qinner}[2]{\langle#1|#2\rangle}

\begin{document}
\title{Characteristic identities for Lie (super)algebras}

\author{Phillip S Isaac$^1$, Jason L Werry and Mark D Gould}

\address{School of Mathematics and Physics, The University of Queensland, St Lucia, QLD 4072,
Australia}

\ead{$^1$psi@maths.uq.edu.au}

\begin{abstract}
We present an overview of characteristic identities for Lie algebras and superalgebras. We outline
methods that employ these characteristic identities to deduce matrix elements of finite dimensional
representations. To demonstrate the theory, we look at the examples of the general linear Lie algebras and Lie
superalgebras.
\end{abstract}

\section{Introduction}

Characteristic identities have proven to be a useful tool for revealing important and explicit information
about representations of Lie algebras and superalgebras. The aim of this article is to give an
overview of how characteristic identities may be used for such purposes. 

It was Dirac in 1936 \cite{Dirac1936} who first employed polynomial identities related to $sl(2)$ in
the context of relativistically invariant wave equations. Later, Lehrer-Ilamed \cite{Lehrer1956}
noted more generally that $n^2$ elements of the enveloping algebra $U(g)$ of a Lie algebra $g$
satisfy $n^2$ identities, which in some cases may be expressed as a single polynomial identity of
degree $n$ for an $n\times n$ matrix with entries from $U(g)$. There were subsequent works, for
example, by Louck \cite{Louck1965}, Mukunda \cite{Mukunda1967} and Louck and Galbraith
\cite{LouckGalbraith1972} in which polynomial identities were encountered for a variety of Lie
algebras. In \cite{BrackenGreen1971,Green1971,Green1975}, Bracken and Green established a general
theory of characteristic identities for classical Lie algebras, which was soon followed by the work
of O'Brien, Cant and Carey \cite{OBCC1977} in which the authors instituted a suitable algebraic
formalism. In a series of papers
\cite{Gould1978v1,Gould1978v2,Gould1980,Gould1981v1,Gould1981v2,Gould1984,Gould1985}, Gould made use
of the characteristic identities to deduce information such as invariants, Wigner coefficients and matrix
elements pertaining to the irreducible representations of the classical Lie algebras. It is also
worth noting that these polynomial identities also extend to finite groups \cite{Gould1987v2} and
quantum groups \cite{GouldZhangBracken1991,GouldLinksBracken1992}.

Following the work of Kac \cite{Kac1977,Kac1978}, Jarvis and Green
\cite{GreenJarvis1979,GreenJarvis1983} investigated characteristic identities and invariants
related to the vector representation of the general linear, special linear and orthosymplectic Lie
superalgebras. Other works soon followed that were related to other Lie superalgebras (e.g.
\cite{JarvisMurray1983}) and more general representations (e.g. \cite{Gould1987}). 
It was not until the more recent work of the current authors \cite{GIW1,GIW2}, however, 
in which the approach of Gould for Lie algebras was generalised to the case of Lie superalgebras,
particularly for the general linear case.

\section{Characteristic identities for Lie algebras}

To demonstrate how the characteristic identities may be utilised, we exhibit the theory in the
context of the Lie algebra $gl(n)$. 

\subsection{Characteristic matrix}

We denote the generators of $gl(n)$ by $a_{ij}$, $i,j=1,\ldots, n$ which satisfy the defining
relations
$$
[a_{ij}, a_{k\ell}] = \delta_{kj}a_{i\ell} - \delta_{i\ell}a_{kj}.
$$
We may assemble the generators into a square $n\times n$ matrix ${\cal A}$ with $(i,j)$ entry $a_{ij}$, i.e.
$$
{\cal A} = 
\left( \begin{array}{cccc}
a_{11} & a_{12} & \cdots & a_{1n}\\
a_{21} & a_{22} & \cdots & a_{2n}\\
\vdots & \vdots &        & \vdots \\
a_{n1} & a_{n2} & \cdots & a_{nn}
\end{array} 
\right)
$$
and define powers of ${\cal A}$ recursively by
$$
\left( {\cal A}^{M+1} \right)_{ij} = \left( {\cal A}^M \right)_{ik}{\cal A}_{kj} 
= {\cal A}_{ik}\left({\cal A}^M\right)_{kj}, \ \ \left( {\cal A}^0 \right)_{ij} = \delta_{ij}.
$$

\subsection{Examples: $gl(1),$ $gl(2)$}

Casimir invariants are defined as $\sigma_1 = a_{ii}$, $\sigma_2 = a_{ij}a_{ji}$, $\sigma_3 =
a_{ij}a_{jk}a_{ki},\ldots,$ $\sigma_M = $tr$\left({\cal A}^M\right).$
To assist the reader in understanding the notation used throughout, consider the following straightforward examples
that were originally presented in \cite{Green1971}. 
For $gl(1)$, (trivial) direct calculation yields 
$$
{\cal A} - \sigma_1=0,
$$ 
which is interpreted to mean ${\cal A}_{ij} - \sigma_1\delta_{ij}=0$ for each $i,j$. 
For $gl(2)$, we may similarly calculate 
$$
{\cal A}^2 - (\sigma_1+1){\cal A} + \frac12\left( \sigma_1^2 + \sigma_1-\sigma_2 \right)=0.
$$
This gives two examples of polynomial identities satisfied by the matrix ${\cal A}$ with
coefficients in the centre of the enveloping algebra of $gl(1)$ and $gl(2)$ respectively.
It is worth noting that in these cases $\sigma_2=\sigma_1^2$, $\sigma_3 = \sigma_1^3$ for $gl(1)$ and
$\sigma_3=\frac32\sigma_1\sigma_2-\frac12\sigma_1^3+\sigma_2-\frac12\sigma_1^2$ for $gl(2)$,
demonstrating that the Casimir invariants so defined are not generally independent.

\subsection{The characteristic identity}

The characteristic identity of ${\cal A}$ for $gl(n)$ on a finite dimensional irreducible
representation $V(\lambda)$ is constructed in the following way. 

In the previous section we saw examples of how the matrix ${\cal A}$ satisfies polynomial identities
with coefficients expressible in terms of the Casimir invariants. If we now look at such polynomial
identities on a finite dimensional irreducible representation $V(\lambda)$, with highest weight 
highest weight $\lambda = (\lambda_1,\lambda_2,\ldots,\lambda_n)$,
we then have from
Schur's Lemma that the Casimir invariants take on constant values on $V(\lambda)$. 
For example, for $gl(n)$, 
$$
\sigma_1=\sum_{j=1}^n\lambda_j,\ \ \ \ \ 
\sigma_2 = \sum_{j=1}^n \lambda_j(\lambda_j+n+1-2j).
$$
Bracken and Green \cite{BrackenGreen1971,Green1971} showed that the identity
$$
\prod_{j=1}^n({\cal A} - \alpha_j)=0
$$
holds on a finite dimensional irreducible representation of $gl(n)$, where $\alpha_j =
\lambda_j+n-r$ are referred to as the characteristic roots.
We highlight the fact that the entries of ${\cal A}$ are now representation matrices.

\subsection{The adjoint matrix}

The matrix ${\cal A}$ is not the only such matrix one may define. Consider taking the  
negative transpose of ${\cal A}$ as
$$
\overline{{\cal A}}_{ij} =  -a_{ji}.
$$
One may also show that 
$$
\prod_{j=1}^n(\overline{{\cal A}} - \overline{\alpha}_j)=0
$$
holds on a finite dimensional irreducible representation of $gl(n)$, where the characteristic roots
are in this case given by $\overline{\alpha}_j = r-1-\lambda_j=n-1-\alpha_j$.

\subsection{General characteristic matrix}

The construction of these characteristic matrices at this stage may seem ad hoc, but there is in
fact a general formalism which relies on the quadratic Casimir element \cite{OBCC1977,Gould1985}. 
Such a matrix can then be constructed for an arbitrary semisimple Lie algebra.

Let $\Delta$ be the co-product of the enveloping algebra $U(g)$ of a semisimple Lie algebra $g$, 
and let $\pi_\mu$ denote any irreducible representation
of $g$ corresponding to module $V(\mu)$. We may define the matrix with algebraic entries
$$
{\cal A}_{\mu} = -\frac12\left[\phantom{\frac12} (\pi_\mu\otimes\id)\Delta(\sigma_2) -
\pi_\mu(\sigma_2)\otimes 1 - I\otimes \sigma_2\phantom{\frac12} \right].
$$
Considering its action on an arbitrary finite dimensional irreducible representation $\pi_\lambda$
gives
$$
{\cal A}^\lambda_{\mu} = -\frac12\left[\phantom{\frac12} (\pi_\mu\otimes\pi_\lambda)\Delta(\sigma_2) -
\pi_\mu(\sigma_2)\otimes I - I\otimes \pi_\lambda(\sigma_2)\phantom{\frac12} \right].
$$
In the case $V(\mu)$ is the vector representation of $g=gl(n)$ (i.e. highest weight $\mu=(1,0,\ldots,0)$) we
obtain $\overline{{\cal A}}$,  and 
for the case we take the dual vector representation of $g=gl(n)$, we obtain ${\cal A}$.

\subsection{Characteristic roots}

For the characteristic roots $\alpha_\nu$ in the general case, first consider the Clebsch-Gordan decomposition
$$
V(\mu)\otimes V(\lambda) = \bigoplus_{\nu} m_{\mu\lambda}^\nu V(\nu).
$$
On each $V(\nu)$, ${\cal A}^\lambda_\mu\in$End$(V(\mu)\otimes V(\lambda))$ takes on the constant value
$$
\alpha_\nu = -\frac12\left[ \phantom{\frac12} \chi_\nu(\sigma_2) - \chi_\mu(\sigma_2) -
\chi_\lambda(\sigma_2) \phantom{\frac12} \right],
$$
where $\chi_\nu(\sigma_2)$ denotes the eigenvalue of $\sigma_2$ on $V(\nu)$, 
$\chi_\lambda(\sigma_2)$ is the eigenvalue of $\sigma_2$ on $V(\lambda)$, and
$\chi_\mu(\sigma_2)$ the eigenvalue of $\sigma_2$ on $V(\mu)$. This results in the characteristic identity
$$
\prod_\nu({\cal A}^\lambda_\mu -\alpha_\nu) = 0.
$$

\subsection{Projection operators}

We may use the characteristic identities to define projection operators, which, as we shall see, turn out to
be a crucial ingredient for determining matrix elements. 

We modify our notation slightly, so that
${\cal A}_n$ is understood to be the characteristic matrix associated with $gl(n)$.
Explicitly, we have from the characteristic identity
$$
P\bmat{n\\ r} = 
\prod_{\ell\neq r} \left( \frac{{\cal A}_n - \alpha_{\ell,n}}{\alpha_{r,n}-\alpha_{\ell,n}} \right),
$$
and
$$
\overline{P}\bmat{n\\ r} = 
\prod_{\ell\neq r} \left( \frac{\overline{{\cal A}}_n -
\overline{\alpha}_{\ell,n}}{\overline{\alpha}_{r,n}-\overline{\alpha}_{\ell,n}} \right).
$$

If we set $V$ to be the vector representation of $gl(n)$, we have the decomposition 
$$
V\otimes V(\lambda) = \bigoplus_{k=1}^n V(\lambda+\Delta_k),
$$
and so
$$
\overline{P}\bmat{n\\ r}:V\otimes V(\lambda)\longrightarrow V(\lambda+\Delta_r)
$$ 
is a projection. Here, $\Delta_k$ is simply an $n$-tuple with a 1 in the $k$th entry and zeroes
elsewhere. 

By contrast, denoting by $V^*$ the dual vector representation of $gl(n)$, we have the decomposition
$$
V^*\otimes V(\lambda) = \bigoplus_{k=1}^n V(\lambda-\Delta_k),
$$
which implies that
$${P}\bmat{n\\ r}:V^*\otimes V(\lambda)\longrightarrow V(\lambda-\Delta_r)
$$ 
is a projection. These projections will be used in what follows. 

\section{Matrix elements of irreducible representations} 

We may utilise the characteristic identity to determine the matrix elements of finite dimensional
irreducible representations. Here we demonstrate the procedure, at the same time highlighting the
power of this approach. Of particular note is the fact that we may deduce, up to a phase factor, not
only the matrix elements of the elementary generators of $gl(n)$, i.e. those of the form $a_{i\ i+1}$, 
but also the nonelementary generators.

\subsection{Vector operators}

Before presenting the details of the matrix element formulae, we first need to describe vector
operators.
Define a $gl(n)$ vector operator as a collection of $n$ operators $\psi_i$ satisfying
$$
[a_{ij},\psi_k] = \delta_{jk}\psi_i.
$$
Related to the dual vector representation, we also define a $gl(n)$ contragredient vector operator
as a collection of $n$ operators $\phi_j$ satisfying
$$
[a_{ij},\phi_k] = -\delta_{ik}\phi_j.
$$
From the work of Green \cite{Green1971}, we have 
$$
\psi_j = \sum_{r=1}^n\psi\bmat{n\\ r}_j,
$$
where
$$\psi\bmat{n\\ r}_j=\psi_i\overline{P}\bmat{n\\ r}_{ij} = P\bmat{n\\ r}_{ji}\psi_i$$ 
are linearly independent and change that value of the highest weight label $\lambda_{r,n}$ by one, leaving the other $\lambda_{k,n}$
unchanged, i.e.
$$
\lambda_{k,n}\psi\bmat{n\\ r}_j = \psi\bmat{n\\ r}_j(\lambda_{k,n} + \delta_{kr}).
$$
Also, in the case of the contragredient vector operator, we have
$$
\phi_j = \sum_{r=1}^n\phi\bmat{n\\ r}_j,
$$
such that
$$  
\lambda_{k,n}\phi\bmat{n \\ r}_j = \phi\bmat{n\\ r}_j(\lambda_{k,n} - \delta_{kr}),
$$
where 
$$
\phi\bmat{n\\ r}_j = \phi_i{P}\bmat{n\\ r}_{ij}=\overline{P}\bmat{n\\ r}_{ji}\phi_i.
$$

\subsection{Branching rules}

The $gl(n)\subset gl(n+1)$ branching rules give the decomposition of a $gl(n+1)$ irreducible
representation into a direct sum of $gl(n)$ irreducible representations:
$$
V(\lambda_{n+1}) = \bigoplus_{\lambda_n}V(\lambda_n), 
$$
where the direct sum is taken over an admissible set of dominant $gl(n)$ weights $\lambda_n$. This
determines the branching rules. Specifically, the highest weights
$$
\lambda_{n+1} = (\lambda_{1,n+1},\lambda_{2,n+1},\ldots,\lambda_{n+1,n+1}),
\ \ \ \lambda_n = (\lambda_{1,n},\lambda_{2,n},\ldots,\lambda_{n,n})
$$
are known to satisfy
$$\lambda_{i,n+1}-\lambda_{i,n}\in\mathbb{Z}_+, 
\ \ \lambda_{i,n} - \lambda_{i+1,n+1} \in \mathbb{Z}_+.$$
By recursion we may deduce the Gelfand-Tsetlin \cite{GT1950} patterns for the entire subalgebra chain
$$
gl(1)\subset gl(2)\subset \cdots\subset gl(n)\subset gl(n+1),
$$ 
which is in one-to-one correspondence with an orthonormal basis of $V(\lambda_{n+1})$.

Now consider the $gl(n+1)$ characteristic matrix:
$$
{\cal A}_{n+1} = 
\left( \begin{array}{cccc|c}
a_{11} & a_{12} & \cdots & a_{1n} & a_{1\ n+1}\\
a_{21} & a_{22} & \cdots & a_{2n} & a_{2\ n+1}\\
\vdots & \vdots &        & \vdots & \vdots\\
a_{n1} & a_{n2} & \cdots & a_{nn} & a_{n\ n+1} \\
\hline
a_{n+1\ 1} & a_{n+1\ 2} & \cdots & a_{n+1\ n} & a_{n+1\ n+1} \\
\end{array} 
\right)
=
\left( \begin{array}{ccc|c} &&&\\ &{\cal A}_n& & \psi\\ &&& \\ 
\hline & \phi &  & a_{n+1\ n+1} \end{array} \right),
$$
where $\psi_j = a_{j\ n+1}$, $\phi_j = a_{n+1\ j}=\psi_j^\dagger$ are $gl(n)$ vector and contragredient vector
operators respectively. The matrix ${\cal A}_{n+1}$ satisfies 
$$
\prod_{j=1}^{n+1}({\cal A}_{n+1} - \alpha_{j,n+1})=0,
$$ 
and the projections are given explicitly by
$$
P\bmat{n+1\\ r}=\prod_{\ell\neq r}^{n+1} \left( \frac{{\cal A}_{n+1} -
\alpha_{\ell,n+1}}{\alpha_{r,n+1}-\alpha_{\ell,n+1}} \right).
$$

\subsection{Construction of invariants}

Following the work of Gould \cite{Gould1981v1}, we are now in a position to construct eigenvalues of invariants
$C_{k,n+1}$ (defined below). 
Consider the $gl(n+1)$ identity expressed in the form 
$$
{\cal A}_{n+1} P\bmat{n+1\\ k} = \alpha_{k,n+1} P\bmat{n+1\\ k}.
$$
Looking at the $(j,n+1)$ entry, $j=1,2,\ldots, n$ gives 
$$
\psi_j C_{k,n+1} = (\alpha_{k,n+1}-{\cal A}_n)_{j\ell} P\bmat{n+1\\ k}_{\ell\ n+1},
$$
where 
$$
C_{k,n+1}=P\bmat{n+1\\ k}_{n+1\ n+1}
$$ 
is a $gl(n)$ invariant. After some manipulation, we arrive at 
$$
P\bmat{n+1\\ k}_{j\ n+1}=\sum_{r=1}^n\psi\bmat{n\\
r}_j(\alpha_{k,n+1}-\alpha_{r,n}-1)^{-1}C_{k,n+1}. 
$$

It is known \cite{Green1975,Gould1978v1} that for any polynomial $p(x)$, 
\begin{align*}
& p({\cal A}_{n+1}) = \sum_{k=1}^{n+1}p(\alpha_{k,n+1})P\bmat{n+1\\ k}\\ 
\Rightarrow &  \delta_{j\ n+1} = \sum_{k=1}^{n+1}P\bmat{n+1\\ k}_{j\ n+1}=0,\\
\Rightarrow & \sum_{r=1}^n\psi\bmat{n\\ r}_j\left( \sum_{k=1}^{n+1}(\alpha_{k,n+1}-\alpha_{r,n}-1)^{-1}C_{k,n+1}
\right)=0 \\
\Rightarrow & \sum_{k=1}^{n+1}(\alpha_{k,n+1}-\alpha_{r,n}-1)^{-1}C_{k,n+1}=0,
\end{align*}
since the $\psi\bmat{n\\ r}_j$ are linearly independent. Also, 
$$
\sum_{k=1}^{n+1}C_{k,n+1}=1,
$$ 
since the $C_{k,n+1}$ are defined in terms of projections. This gives a set of linear equations
that can be solved for $C_{k,n+1}$, giving
$$
C_{k,n+1} = \prod_{p\neq
k}^{n+1}(\alpha_{k,n+1}-\alpha_{p,n+1})^{-1}\prod_{\ell=1}^n(\alpha_{k,n+1}-\alpha_{\ell,n}-1).
$$
Similarly, for 
$$
\overline{C}_{k,n+1}=\overline{P}\bmat{n+1\\k}_{n+1\ n+1},
$$ 
we have
$$
\overline{C}_{k,n+1} = \prod_{p\neq
k}^{n+1}(\alpha_{k,n+1}-\alpha_{p,n+1})^{-1}\prod_{\ell=1}^n(\alpha_{k,n+1}-\alpha_{\ell,n}).
$$

\subsection{Matrix element formula}

In \cite{Gould1978v1}, Gould showed that
\begin{align}
\psi\bmat{n\\ r}\psi^\dagger\bmat{n\\ r}&=\overline{M}_{r,n}P\bmat{n\\ r},\nonumber \\ 
\psi^\dagger\bmat{n\\ r}\psi\bmat{n\\ r} &= M_{r,n}\overline{P}\bmat{n\\ r}, \label{equ1}
\end{align}
where
\begin{align*}
\overline{M}_{r,n} &= (-1)^n\prod_{p=1}^{n+1}(\alpha_{p,n+1}-\alpha_{r,n})\prod_{\ell\neq
r}^n(\alpha_{r,n}-\alpha_{\ell,n}-1)^{-1},\\
M_{r,n} &=  (-1)^n\prod_{p=1}^{n+1}(\alpha_{p,n+1}-\alpha_{r,n}-1)\prod_{\ell\neq
r}^n(\alpha_{r,n}-\alpha_{\ell,n}+1)^{-1}.
\end{align*}
Taking the $(n,n)$ entry of equation (\ref{equ1}) above leads to the following expression for the
action of the elementary generators $a_{n\ n+1}$: 
$$
a_{n\ n+1}\ket{\lambda_{j,k}} = \sum_{r=1}^n\psi\bmat{n\\ r}\ket{\lambda_{j,k}} = \sum_{r=1}^n
N_r^n\ket{\lambda_{j,k} + \Delta_{r,n}},
$$
where
$\ket{\lambda_{j,k}}$ denotes a state corresponding to a Gelfand pattern, 
$\Delta_{r,n}$ denotes a shift in the $r$th highest weight label at the $n$th level, and
\begin{align*}
N^n_r &= \bra{\lambda_{j,k}} M_{r,n}\overline{C}_{r,n}\ket{\lambda_{j,k}}^{1/2}\\ 
\Rightarrow \ \ 
N^n_r &=\left( \frac{(-1)^n\prod_{p=1}^{n+1}(\lambda_{p,n+1} -
\lambda_{r,n}+r-p)\prod_{\ell=1}^{n-1}(\lambda_{r,n}-\lambda_{\ell,n-1}+\ell-r+1)}
{\prod_{\ell\neq
r}^n(\lambda_{r,n}-\lambda_{\ell,n}+\ell-r)(\lambda_{r,n}-\lambda_{\ell,n}+\ell-r-1)} \right)^{1/2}.
\end{align*}
Similar expressions may be obtained for $a_{n+1\ n}$.

Using similar techniques, we may also determine matrix element formulae for $a_{\ell\ n+1}$:
$$
a_{\ell\ n+1}\ket{\lambda_{j,k}} = \sum_{i_n=1}^n\sum_{i_{n-1}=1}^{n-1}\cdots\sum_{i_\ell=1}^{\ell}
N\bmat{n&\cdots&\ell\\i_n&\cdots&i_\ell}\ket{\lambda_{j,k} + \Delta_{i_n,n}+\cdots
+\Delta_{i_\ell,\ell}},
$$
where
\begin{equation}
N\bmat{n&\cdots&\ell\\i_n&\cdots&i_\ell} = \pm\prod_{r=\ell}^nN^r_{i_r}\prod_{r=\ell+1}^n\left[
(\lambda_{i_r,r}-\lambda_{i_{r-1},r-1}+i_{r-1}-i_r)
(\lambda_{i_r,r} - \lambda_{i_{r-1},r-1}+i_{r-1}-i_r+1)\right]^{-1/2}.
\label{mels}
\end{equation}
Similarly for $a_{n+1\ \ell}$.

\section{Generalisation to Lie superalgebras}

For the Lie superalgebra $gl(m|n)$, we now have generators $a_{pq}$, $p,q=1,2,\ldots,m+n$ satisfying
$$
[a_{pq},a_{rs}] \equiv a_{pq}a_{rs} - (-1)^{[(p)+(q)][(r)+(s)]}a_{rs}a_{pq}
= \delta_{qr}a_{ps} - (-1)^{[(p)+(q)][(r)+(s)]}\delta_{ps}a_{rq}.
$$
where
$$
(p) =  \left\{ \begin{array}{rl} 0, & p=1,\ldots,m, \\ 1, & p=m+1,\ldots, m+n. \end{array} \right.
$$
Analogous to the Lie algebra case, we define the characteristic matrices by setting 
$$
{\cal A}_{pq}=-(1)^{(p)} a_{pq},\ \ \ \overline{{\cal A}}_{pq} = -(-1)^{(p)(q)}a_{qp}.
$$
On the irreducible representation $V(\Lambda)$, with 
$\Lambda = (\Lambda_1,\Lambda_2,\ldots,\Lambda_m|\Lambda_{m+1},\ldots,\Lambda_{m+n})$, it is also
possible to show \cite{Gould1987} that the following characteristic identities hold
$$
\prod_{p=1}^{m+n}( {\cal A} - \alpha_p)=0,\ \ \ \prod_{p=1}^{m+n}( \overline{{\cal A}} -
\overline{\alpha}_p)=0,
$$
with associated characteristic roots
$$
\alpha_p = (-1)^{(p)}(\Lambda_p+m-p)-n,\ \ \ \overline{\alpha}_p = m-(-1)^{(p)}(\Lambda_p+m+1-p).
$$
By analogy with the methods outlined for $gl(n)$, we may construct similar objects for $gl(m|n)$
as listed below.

\begin{itemize}
\item[1.] Projection operators $\displaystyle{P\bmat{m+n\\ r}}$;
\item[2.] Vector operators $\displaystyle{\psi_r =(-1)^{(r)}a_{r\ m+n+1}}$;
\item[3.] Branching rules;
\item[4.] Invariants  ($C_{r,m+n}$, $\overline{C}_{r,m+n}$, $M_{r,m+n}$ and
$\overline{M}_{r,m+n}$);
\item[5.] Matrix elements. 
\end{itemize}

There is, however, one major complication in the context of Lie superalgebras. The representation
theory is not as well-understood as that of Lie algebras, and in particular the branching rules for
an arbitrary irreducible representation are not known. In fact, in general it is not even known if
an irreducible representation is completely reducible, i.e. there is no analogue of Weyl's Theorem
in general. To avoid this problem, we look at a special subset of the irreducible representations
first considered by Scheunert, Nahm and Rittenberg \cite{SNR1977}.

\subsection{On complete reducibility}

The so-called star representations were first considered by Scheunert, Nahm and Rittenberg
\cite{SNR1977} in order to ensure complete reducibility.
Let $\qinner{\ }{\ }$ be a well-defined, positive-definite, invariant, sesquilinear form defined on
an irreducible highest weight module $V(\Lambda)$ of $gl(m|n)$:
$$
\qinner{a_{pq}v}{w} = (-1)^{\epsilon[(p)+(q)]}\qinner{v}{a_{qp}w},
\ \ \ 
\epsilon = 0,1.
$$ 

For the case $\epsilon=0$, $V(\Lambda)$ is said to be a type 1 star irrep. Otherwise, for 
$\epsilon=1$, $V(\Lambda)$ is called a type 2 star irrep. We have complete reducibility in these
cases, and so determine branching rules, which is the one complication in the above procedure.

\subsection{Classification of unitary (star) irreducible representations}

The star representations, which we now refer to as ``unitary'' due to the existence of the
positive-definite form (i.e. an inner product), were classified by Gould and Zhang
\cite{GouZha1990v1,GouZha1990v2}. Here we only focus on the type 1 case. 

In particular, $V(\Lambda)$, with highest weight
$\Lambda=(\lambda_1,\lambda_2,\ldots,\lambda_m|\lambda_{\bar{1}},\ldots,\lambda_{\bar{n}})$ is a
type 1 star irrep. iff
\begin{itemize}
\item[(i)] $\lambda_m+\lambda_{\bar{n}}>n-1$ (typical), or
\item[(ii)] there exists an odd index $\mu=\{ \bar{1},\ldots,\bar{n} \}$ such that
$\lambda_m+\lambda_\mu+1-\mu=0 = \lambda_{\bar{n}} - \lambda_\mu$ (atypical).
\end{itemize}
There are similar conditions for type 2 star irreps which may be found in \cite{GouZha1990v1}, but
we will not include the details here.

As indicated in the paper \cite{GIW2}, we set 
$$
\varepsilon = (1,1,\ldots, 1|0,0,\ldots,0),\ \ \delta = (-1,-1,\ldots,-1|1,1,\ldots,1).
$$ 
Then a type 1 star irrep. $V(\Lambda)$ of $gl(m|n)$ has highest weight expressible in the form
\begin{equation}
\Lambda = \Lambda_0 + \gamma \varepsilon + \omega \delta,
\label{hwform}
\end{equation}
where
$\Lambda_0$ corresponds to a covariant tensor representation, i.e. it occurs within a finite tensor
product of the vector module $V(1,\dot{0}|\dot{0})$ with itself,
$\gamma=0,1,2,\ldots, n-1$ or $n-1<\gamma\in\mathbb{R},$ and $\omega\in\mathbb{R}.$

\subsection{Form of the matrix elements}

The main outcome of our research so far is that for type 1 star irreps., generators of $gl(m|n+1)$ have action of the form
$$
a_{\ell\ p+1}\ket{\Lambda_{q,s}} = \sum_{u} N[u_p,u_{p-1},\ldots u_{\ell +1},
u_{\ell}]\ket{\lambda_{q,s} + \Delta_{u_p,p} + \cdots +\Delta_{u_\ell,\ell}}, \ \ \ell<p+1,
$$ 
and similarly for $a_{p+1\ \ell}$. The precise form of the coefficients $N[u_p,u_{p-1},\ldots u_{\ell +1},
u_{\ell}]$ (i.e. matrix elements) will not be given here, but are similar in form to the matrix
elements of equation (\ref{mels}) above which are related to the Lie algebra $gl(n)$. In fact, for the cases under
consideration, we make the point that the procedure closely follows that already presented for
$gl(n)$. For the fully explicit details of matrix element formulae and other features such as
branching rules related to the type 1 unitary irreps of $gl(m|n)$, see \cite{GIW2}. 

Based on the convenient form of the highest weight given by equation (\ref{hwform}), we see that
our results agree with those presented in the work of Stoilova and van der Jeugt \cite{StoiVan2010},
Molev \cite{Molev2011} and also the essentially typical representations
considered by Palev \cite{Palev1987,Palev1989}.

\section{Future work}

Currently work is in preparation for providing the explicit details of the type 2 unitary irreps.
It is also of interest to consider how to treat the cases of representations that do not admit an
invariant inner product, and hence are not guaranteed to be completely reducible. One possibility is to
investigate mixed tensor representations. That is, those representations that occur within the
tensor product of a type 1 unitary irrep. with a type 2 unitary irrep. 

Having made some progress for the classical Lie superalgebras, it is also of interest to determine the
analogous matrix element formulae for the quantum group counterparts.

Finally, our procedure may be considered beyond the vector representation (a type 1 unitary irrep.)
and the dual vector representation (a type 2 unitary irrep.), leading to a more general pattern
calculus. This remains the topic of future work.

\ack 

M.D.G. and P.S.I are supported by the Australian Research Council through Discovery Project
DP140101492. J.L.W. acknowledges the support of an Australian Postgraduate Award.


\section*{References}
\begin{thebibliography}{9}
\bibitem{Dirac1936} Dirac P A M 1936 {\it Proc. R. Soc. Lond. A} {\bf 155} 447.
\bibitem{Lehrer1956} Lehrer Y 1956 {\it Bull. Res. Counc. Israel} {\bf 5A} 197.
\bibitem{Louck1965} Louck J D 1965 {\it J. Math. Phys.} {\bf 6} 1786.
\bibitem{Mukunda1967} Mukunda N 1967 {\it J. Math. Phys.} {\bf 8} 1069.
\bibitem{LouckGalbraith1972} Louck J D and Galbraith H W 1972 {\it Rev. Mod. Phys.} {\bf 44} 540.
\bibitem{BrackenGreen1971} Bracken A J and Green H S 1971 {\it J. Math. Phys.} {\bf 12} 2099.
\bibitem{Green1971} Green H S 1971 {\it J. Math. Phys.} {\bf 12} 2106.
\bibitem{Green1975} Green H S 1975 {\it J. Austral. Math. Soc. (Series B)} {\bf 19} 129.
\bibitem{OBCC1977} O'Brien D M, Cant A and Carey A L 1977 {\it Ann. Inst. Henri
Poincar\'e, Section A: Physique th\'eorique} {\bf 26} 405.
\bibitem{Gould1978v1} Gould M D 1978 {\it J. Austral. Math. Soc. (Series B)} {\bf 20} 290.
\bibitem{Gould1978v2} Gould M D 1978 {\it J. Austral. Math. Soc. (Series B)} {\bf 20} 401.
\bibitem{Gould1980} Gould M D 1980 {\it J. Math. Phys.} {\bf 21} 444.
\bibitem{Gould1981v1} Gould M D 1981 {\it J. Math. Phys.} {\bf 22} 15.
\bibitem{Gould1981v2} Gould M D 1981 {\it J. Math. Phys.} {\bf 22} 2376.
\bibitem{Gould1984} Gould M D 1984 {\it J. Phys. A: Math. Gen.}  {\bf 17} 1.
\bibitem{Gould1985} Gould M D 1985 {\it J. Austral. Math. Soc. (Series B)} {\bf 26} 257.
\bibitem{Gould1987v2} Gould M D 1987 {\it J. Phys. A: Math. Gen.} {\bf 20} 2657.
\bibitem{GouldZhangBracken1991} Gould M D, Zhang R B and Bracken A J 1991 {\it J. Math. Phys.} {\bf
32} 2298.
\bibitem{GouldLinksBracken1992} Gould M D, Links J and Bracken A J 1992 {\it J. Math. Phys.} {\bf
33} 1008.
\bibitem{Kac1977} Kac V G 1977 {\it Adv. in Math.} {\bf 26} 8.
\bibitem{Kac1978} Kac V G 1978 {\it Lecture Notes in Math.} {\bf 676}, Springer, Berlin, 597.
\bibitem{GreenJarvis1979} Jarvis P D and Green H S 1979 {\it J. Math. Phys.} {\bf 20} 2115.
\bibitem{GreenJarvis1983} Green H S and Jarvis P D 1983 {\it J. Math. Phys.} {\bf 24} 1681.
\bibitem{JarvisMurray1983} Jarvis P D and Murray M K 1983 {\it J. Math. Phys.} {\bf 24} 1705.
\bibitem{Gould1987} Gould M D 1987 {\it J. Austral. Math. Soc. (Series B)} {\bf 28} 310.
\bibitem{GIW1} Gould M D, Isaac P S and Werry J L 2013 {\it J. Math. Phys.} {\bf 54} 013505.
\bibitem{GIW2} Gould M D, Isaac P S and Werry J L 2014 {\it J. Math. Phys.} {\bf 55} 011703.
\bibitem{GT1950}
Gelfand I M and Tsetlin M L 1950 {\it Dokl. Akad. Nauk., SSSR} {\bf 71} 825 (Russian).
English translation in: Gelfand I M 1988 ``Collected Papers'', Vol II, Berlin:
Springer-Verlag 653.
\bibitem{SNR1977} Scheunert M, Nahm W and Rittenberg V 1977 {\it J. Math. Phys.} {\bf 18} 146.
\bibitem{GouZha1990v1} Gould M D and Zhang R B 1990 {\it J. Math. Phys.} {\bf 31} 2552.
\bibitem{GouZha1990v2} Gould M D and Zhang R B 1990 {\it Lett. Math. Phys.} {\bf 20} 221.
\bibitem{StoiVan2010} Stoilova N I and Van der Jeugt J 2010 {\it J. Math. Phys.} {\bf 51} 093523.
\bibitem{Molev2011} Molev A I 2011 {\it Bull. Inst. Math. Acad. Sinica} {\bf 6} 415.
\bibitem{Palev1987} Palev T D 1987 {\it Funct. Anal. Appl.} {\bf 21} 245.
\bibitem{Palev1989} Palev T D 1989 {\it Funct. Anal. Appl.} {\bf 23} 141.
\end{thebibliography}
\end{document}